# Pressure-Induced Phase Transition Versus Amorphization in Hybrid Methylammonium Lead Bromide Perovskite

Akun Liang,[1,2] Robin Turnbull,[1] Catalin Popescu,[3] Ismael Fernandez-Guillen,[4] Rafael Abargues,[4] Pablo P. Boix,[4] and Daniel Errandonea[1,*]

[1]Departamento de Física Aplicada-ICMUV-MALTA Consolider Team, Universitat de València, c/Dr. Moliner 50, 46100 Burjassot (Valencia), Spain

[2]Centre for Science at Extreme Conditions and School of Physics and Astronomy, University of Edinburgh, Edinburgh EH9 3FD, United Kingdom

[3]CELLS-ALBA Synchrotron Light Facility, Cerdanyola, 08290 Barcelona, Spain

[4]Institut de Ciència dels Materials, Universidad de Valencia, C/J. Beltran 2, 46980 Paterna, Spain

*Corresponding author: daniel.errandonea@uv.es

**TOC IMAGE**

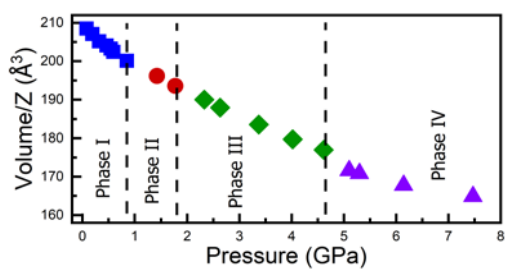

KEYWORDS: hybrid perovskite, phase transition, powder XRD, electronic properties

ABSTRACT

The crystal structure of $CH_3NH_3PbBr_3$ perovskite has been investigated under high-pressure conditions by synchrotron-based powder X-ray diffraction. We found that after the previously reported phase transitions in $CH_3NH_3PbBr_3$ ($Pm\bar{3}m{\rightarrow}Im\bar{3}{\rightarrow}Pmn2_1$), which occur below 2 GPa, there is a third transition to a crystalline phase at 4.6 GPa. This transition is reported here for the first time contradicting previous studies which reported amorphization of $CH_3NH_3PbBr_3$ between 2.3 and 4.6 GPa. Our X-ray diffraction measurements show that $CH_3NH_3PbBr_3$ remains crystalline up to at least 7.6 GPa, the highest pressure covered by experiments. The new high-pressure phase is also described by the space group $Pmn2_1$, however the transition involves abrupt changes in the unit-cell parameters and a 3% decrease of the unit-cell volume. Our conclusions are confirmed by optical-absorption experiments, visual observations and by the fact that pressure-induced changes up to 10 GPa are reversible. The optical studies also allow for the determination of the pressure dependence of the band-gap energy which is discussed using the structural information obtained from X-ray diffraction.

## 1. Introduction

Over the last decades, hybrid organic–inorganic perovskites (HOIPs) have experienced a stunning development because they are outstanding light-absorbing materials for highly efficient photovoltaic cells.[1] Amongst HOIPs, methylammonium (MA, $CH_3NH_3$) lead halide perovskites, such as $MAPbBr_3$ and $MAPbI_3$, have received the most attention and perovskite solar cells have consequently achieved a photovoltaic efficiency of 25%.[2] The crystal structure of these compounds can be described as an inorganic sub-lattice of corner-sharing $PbBr_6$ ($PbI_6$) octahedral units hosting the MA molecules.[3] Because of the high compressibility of the inorganic host lattice and the organic molecule, the crystal structure of hybrid perovskites can be easily modified applying an external pressure.[1] In the case of $MAPbBr_3$ two phase transitions have been reported below 4.6 GPa.[4-8] Pressure not only modifies the crystal structure, but also the electronic properties, including the band-gap energy, thereby becoming a powerful tool to strengthen the current understanding of the properties of $MAPbBr_3$ as well as other HOIPs, and to provide information relevant for optimizing the photovoltaic performance.[1] Regarding the crystal structure, a recent single-crystal X-ray diffraction study revealed that $MAPbBr_3$ undergoes two phase transitions following the space-group sequence: $Pm\bar{3}m \rightarrow Im\bar{3} \rightarrow Pmn2_1$.[4] The phase transitions occur at 0.8 and 1.8 GPa, respectively. Ref. 4 unveiled the occurrence of the pressure-induced nonpolar/polar transition ($Im\bar{3} \rightarrow Pmn2_1$) in $MAPbBr_3$ for the first time. Amorphization has also been reported to take place in $MAPbBr_3$ at higher pressures between 2.5 and 4 GPa, and it has been proposed that the difference in amorphization pressure relates to the degree of non-hydrostatic stress in the experiments.[5-7] In the case of $MAPbI_3$ a gradual amorphization was believed to occur at 2.7 GPa.[9] However, it was more recently shown that amorphization does not occur in $MAPbI_3$ up to 10.6 GPa, with a phase transition to a crystalline phase occurring at lower pressures wherein a degree of disorder is introduced into the perovskite lattice by a pressure-induced freeze-out of the MA cation motion.[10] This concept of statistical disorder was confirmed by Raman, infrared, and X-ray absorption spectroscopy.[11] Interestingly, the transition reported at 2.7 GPa in MAPbI$_3$ induces an abrupt increase of the band-gap energy. In this work, the pressure-induced amorphization of MAPbBr$_3$ has been re-examined by powder X-ray diffraction (XRD) up to 7.6 GPa. Instead of amorphization, we found a first-order isostructural phase transition at 4.6 GPa. The transition favors a smearing of the XRD patterns, but not the loss of crystallinity. These conclusions are supported by optical-absorption measurements and visual observations which do not show any evidence of amorphization up to 10.6 GPa. The newly reported transition triggers an abrupt increase of 0.27 eV in the band-gap energy.

## 2. Methods

PbBr$_2$ (98% purity, Fisher Chemical) and MABr (98% purity, Ossila) were dissolved in Dimethylformamide (DMF, 99.8% purity, Sigma Aldrich). The solution was stirred at ambient conditions until the precursors were dissolved. The product was then filtered with 0.2 mm pore size filter, kept in a closed vial of 20 cm$^3$, and heated up to 80 °C in an oil bath. Then, it was kept at 80 °C for 24 hours. Reproducible size crystals are obtained with this method. A fine powder used for XRD measurements was obtained by grinding the resulting single-crystal sample. Powder XRD experiments were carried out at the BL04-MSPD beamline of ALBA-CELLS synchrotron.[12] A membrane-type diamond anvil cell (DAC), with culets of 400 μm in diameter, was used to generate high-pressure. A hole with a diameter of 200 μm drilled in the center of a pre-indented stainless-steel gasket was used to contain the finely ground powder obtained from the single crystals of the sample. Silicone oil was used as pressure-transmitting medium in order to minimize the possibility of chemical reaction with the sample,[13] and the ruby fluorescence method was used for pressure calibration.[14] In the pressure range covered by the experiments silicone oil can be considered as a quasi-hydrostatic pressure medium.[13] The wavelength of the monochromatic X-ray beam was 0.4642 Å, and the spot size of the X-ray was 20 × 20 μm (full width at half maximum). A Rayonix SX165 CCD image plate was used to collect the diffraction patterns. The two-dimensional diffraction images were integrated to conventional XRD patterns using DIOPTAS.[15] The FullProf[16] suit was used to perform Rietveld refinements.[17] The protocol for refinement was as follows. The background of the XRD pattern was fitted with a Chebyshev polynomial function with six parameters, the shape of Bragg peaks was modeled with a pseudo-Voigt function. The atomic positions and occupancy factors were fixed to the values determined previously from single-crystal XRD experiments.[4] Unit-cell parameters and the overall displacement factor were assumed as fitting parameters. Optical-absorption experiments were performed using the same high-pressure device. In this case, we used small single crystals, cut with a sharp knife from one as-grown crystal, of about $100 \times 100$ μm$^2$ in size and 10 μm in thickness. The sample-in and sample-out method was used to acquire the optical absorption spectra using the set-up and methods described in our previous work.[18, 19]

### 3. Results and discussion

Powder XRD patterns of MAPbBr$_3$ at selected pressures are shown in **Figure 1**. The patterns up to 4.6 GPa have been examined in detail in our previous report.[4] In brief, at pressures lower than 0.9 GPa, the data can be well indexed by the ambient-pressure cubic crystal structure (Phase I, space group: $Pm\bar{3}m$). Above this pressure, and up to 1.8 GPa, XRD patterns can be assigned to the cubic high-pressure structure (Phase II, space group: $Im\bar{3}$). Above 1.8 GPa, and up to 4.6 GPa, the XRD data can be satisfactorily explained by the recently proposed high-pressure orthorhombic structure (Phase III, space group: $Pmn2_1$). To support this statement, we include Rietveld refinements at 0.1, 1.4, and 2.6 GPa in **Figure 1**. The goodness-of-fit-parameters obtained are: $R_p$ = 1.04%, $R_{wp}$ = 1.42%, and $\chi^2$ = 1.02 at 0.1 GPa; $R_p$ = 1.17%, $R_{wp}$ = 1.51%, and $\chi^2$ = 1.04 at 1.4 GPa; and $R_p$ = 1.80%, $R_{wp}$ = 2.71%, and $\chi^2$ = 1.13 at 2.6 GPa. At the pressure of 4.6 GPa the observed peaks can also be assigned to the same orthorhombic structure. However, a notable decrease of the intensity of the diffraction peaks is observed. The smearing of the signal is even more significant at 5.1 GPa. At this pressure, there is also the merging of some of the Bragg peaks. Despite the loss of intensity in the XRD pattern, at 5.1 GPa and above, the peaks are sharp enough to support the existence of a crystalline phase with a long-range order (Phase IV). In addition, we do not notice any relevant change in the background or the appearance of the typical amorphous halo[20, 21] in the diffraction patterns up to 7.6 GPa.

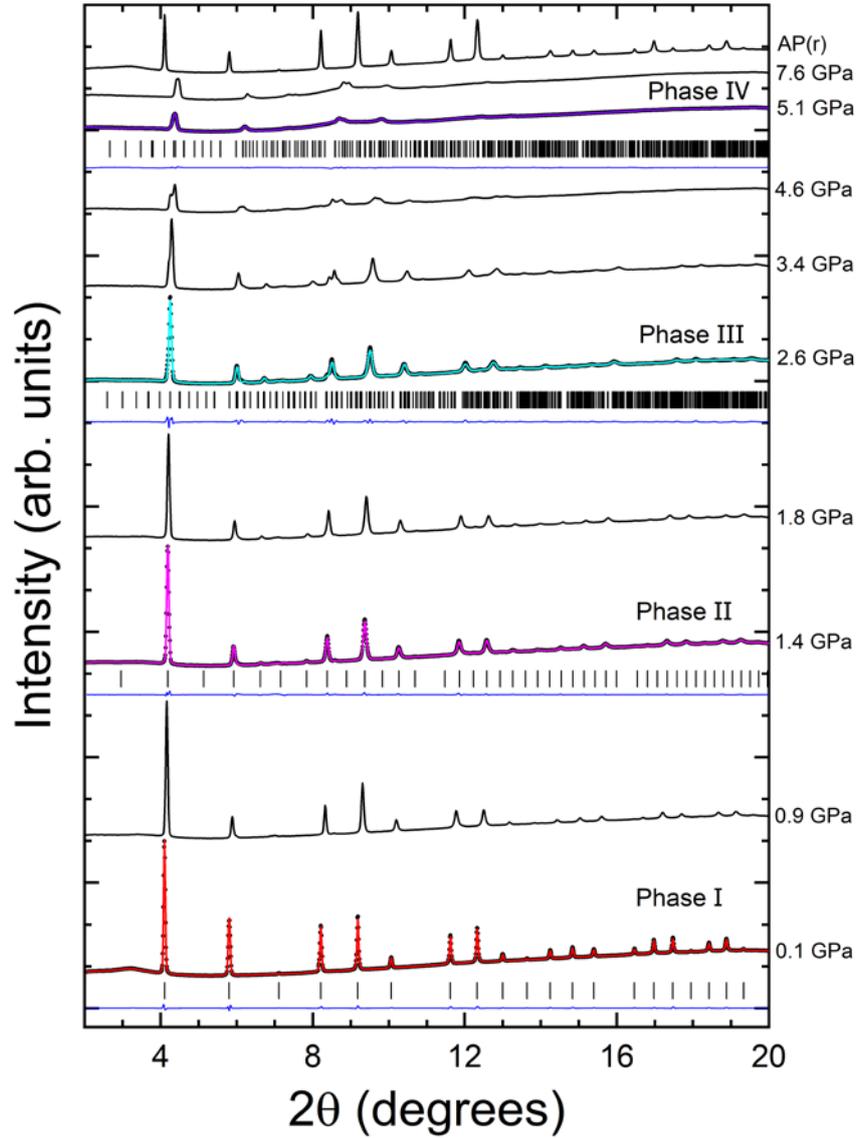

**Figure. 1.** Integrated high-pressure powder XRD patterns from experiments on MAPbBr$_3$. The pressure of each pattern is indicated. AP(r) identifies the pattern measured at ambient pressure (AP) after decompression (recovery, r). At 0.1 GPa the experimental data are shown with black circles, the Rietveld profile with a red line, the residuals with a blue line and reflection positions with vertical ticks. At 1.4 GPa the experiment is shown with black circles, the Rietveld profile with a magenta line, the residuals with a blue line and reflection positions with vertical ticks. At 2.6 GPa the experiment is shown with black circles, the Rietveld profile with a cyan line, the residuals with a blue line and reflection positions with vertical ticks. At 5.1 GPa the experiment is shown with black circles, the Rietveld profile with a purple line, the residuals with a blue line and reflection positions with vertical ticks.

The changes induced in the diffraction patterns at 5.1 GPa do not support amorphization and are more consistent with a gradual disordering of the crystal structure[22] introduced at different length scales in the perovskite lattice by the pressure-induced freeze-out of the MA cation motion as it occurs in MAPbI$_3$.[10] Our conclusions are also in line with the study from Yesudhas *et al.*[23] where the onset of amorphization is set above 5.8 GPa according to Raman experiments. They are also in agreement with recent Raman and photoluminescence experiments that propose a third crystalline-crystalline phase transition and show that the fourth crystalline phase remains crystalline up to circa 7 GPa rather than becoming amourphous.[24] Furthermore, the pressure-induced structural changes observed in powder XRD patterns of MAPbBr$_3$ are totally reversible upon decompression (see **Figure 1**), which is unusual for pressure-induced amorphization.[25] Commonly, when releasing pressure after amorphization, the characteristic halo of a non-crystalline material is observed in addition to weak Bragg reflections of the stable polymorph.[25] This is because pressure-induced amorphization is a first-order transition with dramatic changes in the structure of the solids.

For the crystal structure of the new high-pressure phase (Phase IV, purple in **Figure 1**), we have found that a crystal structure described by the same space group as the orthorhombic phase III (cyan in **Figure 1**, *Pmn*2$_1$) can account for the peaks in the XRD patterns measured from 5.1 to 7.6 GPa. This conclusion is supported by a Rietveld refinement shown in **Figure 1**. In this case, the goodness-of-fit-parameters obtained are R$_p$ = 2.24%, R$_{wp}$ = 3.34%, and $\chi^2$ = 1.02. The unit-cell parameters at 5.1 GPa are $a$ = 11.211(5) Å, $b$ = 15.592(7) Å, and $c$ = 7.860(3) Å. However, the atomic positions, which were chosen based on the structure of Phase III, were not refined. Single-crystal XRD measurement would be probably needed to unambiguously determine the structural model of Phase IV. A thermal annealing of the sample, to suppress/reduce residual stress after three successive phase transition, would be beneficial to acquire good quality single-crystal XRD data of Phase IV. According to these parameters there is an abrupt change in $a$, $b$, and $c$ at the transition and a 3% reduction in the unit-cell volume. Therefore, the transition can be described as a first-order isostructural transition. Interestingly, $b \simeq \sqrt{2}a \simeq 2c$. This makes the new high-pressure structure more symmetric, which is reflected by the merging of Bragg peaks, for instance, the most intense peaks at low angles. The change of unit-cell parameters at the transition will necessarily modify the tilting of PbBr$_6$ octahedra, reducing the size of cavities containig the MA molecules, which can spatially lock MA molecules in different directions. This local dissorder will affect the intensity of XRD patterns. However, it does not destroy the long-range order as would be needed to trigger amorphization.

From powder XRD patterns, we have extracted the pressure dependence of unit-cell parameters. The results are presented in **Figure 2**. In the figure, vertical dashed lines indicate the transition pressure. However, we cannot rule out the phase co-existence between successive phases, at least in such a short pressure range, due to the pressure steps used in the experiment. In **Figure 2a** it can be seen that there is a discontinuity in *a*, *b*, and *c* beyond 4.6 GPa. The discontinuity in the volume is more noticeable (see **Figure 2b**). In the new high-pressure phase, there is also a decrease of the compressibility. By fitting the four data points we measured with a second-order Birch-Murnaghan equation of state (EOS),[26] the zero-pressure bulk modulus of Phase IV is determined to be 33(6) GPa, which is 70% larger than the same parameter for the three phases measured below 4.6 GPa.[4] A similar reduction of compressibility has been previously reported for highly compressible halide compounds when first-order isostructural transitions occur.[27]

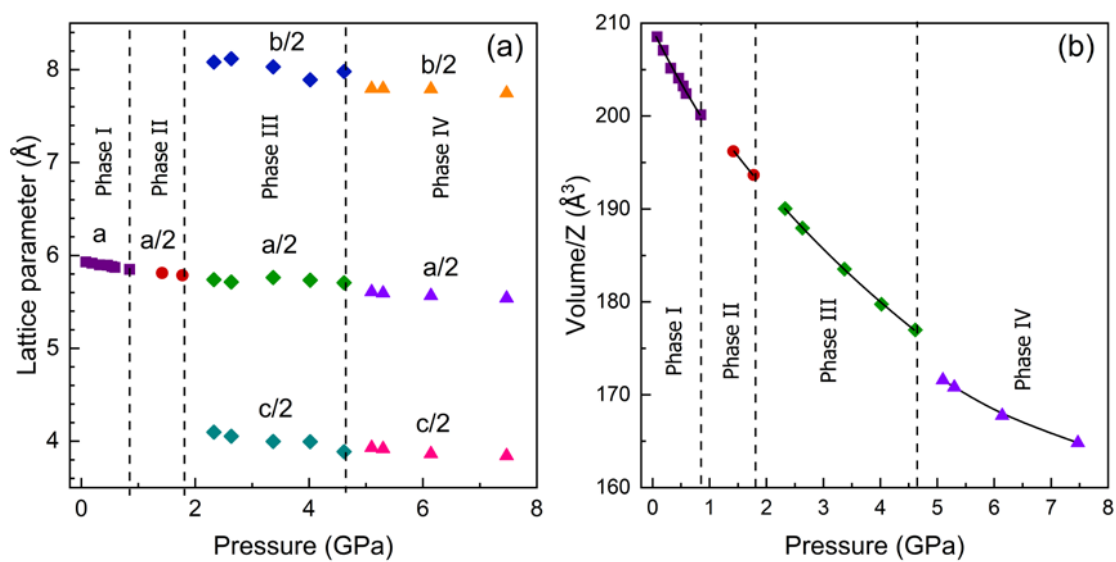

**Figure. 2.** Pressure dependence of the unit-cell parameters (a) and unit-cell volume (b) of MAPbBr$_3$. Different phases and parameters are shown with different colors. The vertical dashed lines indicate the phase transition pressures. In (b) the black solid lines are the second-order Birch-Murnaghan fits, reported here for the Phase IV (purple triangles) and in Ref. 4 for the other three phases.

In addition to the XRD experiments, we have carried out a series of optical studies. In **Figure 3** we show a collection of images taken under a microscope of a MAPbBr$_3$ crystal loaded in a diamond-anvil cell at different pressures. Images were captured under compression and decompression. Before the first phase transition (Phase I), the color of the sample gradually changes moving from a light orange color towards dark orange, indicating a red-shift of the band-gap energy. After the transition, and up to 4.1 GPa, (*i.e.* in the first two high-pressure phases, Phases II and III) the sample gradually becomes light yellow indicating a blue-shift of the band-gap energy. From 4.1 to 5.2 (*i.e.* in Phase IV) there is an abrupt color change that corresponds to a strong blue shift of the band-gap energy. In the micrographs shown in **Figure 3** it can be seen that there are no obvious changes in the shape or crystallinity of the sample, which is not expected if amorphization occurs between 2.5 and 4 GPa as previously reported.[5-7] Usually, amorphization is characterized by the developing of cracks and other microstructural features in the crystal which can be visually observed under the microscope and are related to the creation of dislocation densities which are precursors of the crystalline-to-amorphous transformation.[28, 29] None of these features are observed in MAPbBr$_3$ in our experiments. Upon decompression, the changes in the color of the crystal are fully reversible.

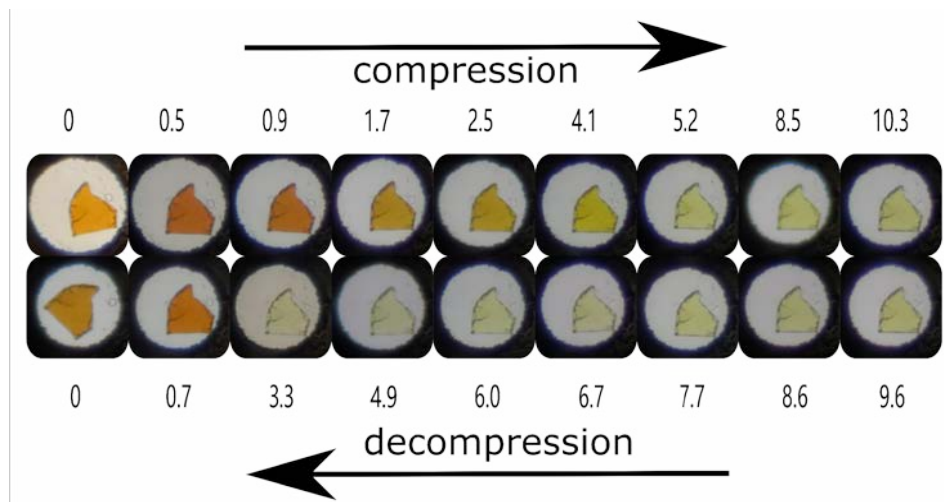

**Figure. 3.** Photomicrographs of a MAPbBr$_3$ single crystal in the sample chamber at selected pressures taken during compression up to 10.3 GPa (top) and decompression (bottom). Color changes are caused by the modification of the band-gap energy. Pressures are indicated in GPa.

Three independent high-pressure optical-absorption experiments have been carried out to determine the pressure dependence of the band-gap energy of MAPbBr$_3$ up to 10 GPa. The optical absorption spectra of one of the experiments (exp 2) is shown in **Figure 4a**. The absorption spectra have a sharp absorption associated with the fundamental band gap. Above 4.6 GPa (phase IV), at energies smaller than the optical gap, we did not observe the residual optical absorption typical of amorphous materials.[30] This supports the conclusions we made from powder XRD experiments. From the optical-absorption experiments, we obtained the pressure-dependence of the band-gap energy summarized in **Figure 4b**. As observed in **Figures 3 and 4**, the band-gap energy in Phase I continuously red-shifts from ambient pressure until the occurrence of the first phase transition. After the transition, the band-gap energy blue-shifts under increasing pressure up to 4.6 GPa. At the phase transition pressure between Phase III and IV, there is an abrupt increase of the band-gap energy of approximately 0.27 eV. Such a change supports the occurrence of a phase transition as determined from XRD experiments at the same pressure. From 4.6 to 10.6 GPa the band-gap energy remains nearly constant. Consequently, the behavior of the band-gap energy supports the argument that there is no other phase transition up to 10.6 GPa. A similar increase of the band-gap energy has been reported for MAPbI$_3$ at 3 GPa.[10] However, the phase transition was originally misinterpreted as the pressure-induced amorphization. The analogous behavior of the two lead halide hybrid perovskites provides additional support to our interpretation of the pressure-induced crystal structural changes happening at 4.6 GPa in MAPbBr$_3$.

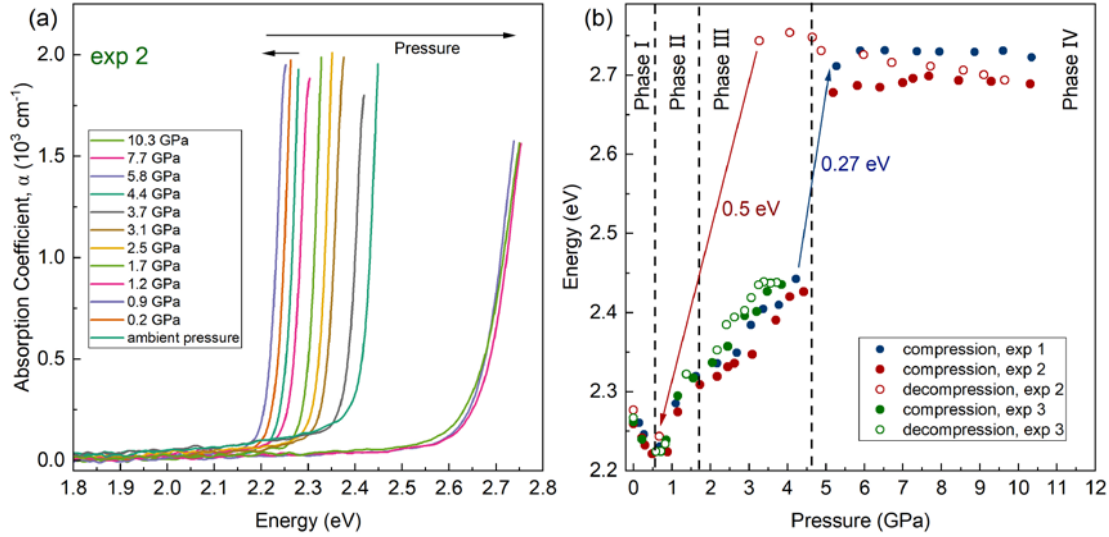

**Figure. 4.** (a). Optical absorption spectra of MAPbBr$_3$ at selected pressures from one of the three independent experiments (exp 2). (b). Pressure dependence of band-gap energy of MAPbBr$_3$ obtained from three independent different experiments shown in blue (exp 1), red (exp 2), and green (exp 3). Solid circles are from experiments performed on compression and empty symbols from experiments performed on decompression.

In two of the optical-absorption experiments we acquired measurements on sample decompression. In one of these experiments the maximum pressure is 3.8 GPa (exp 3, green circles in **Figure 4b**); a pressure below that of third phase-transition pressure. In this experiment (green), the observed changes are fully reversible upon decompression. In the second experiment (exp 2, red circles) which includes data acquired upon decompression, the maximum pressure achieved was 10.6 GPa; a pressure which exceeds the phase transition pressure of 4.6 GPa. In this case, the band gap indicates that the sample remains in phase IV under pressure release down to 3 GPa. The subsequent data point has been measured at 0.7 GPa corresponding to phase II. In contrast with compression, during decompression the pressure increments cannot be fine-tuned with the same precision. This is why there is a 2 GPa step between subsequent measurements on decompression below 3 GPa. The value of the band gap agrees with the value determined at the same pressure during compression. At ambient pressure, after the pressure is fully released, we measured in both cases (green and red) the same band-gap energy as before compression. This shows that the changes induced by pressure in the electronic band structure are fully reversible, which agrees with the reversibility observed in the structural changes.

Knowing that the valence band maximum of MAPbBr$_3$ at ambient pressure is mainly contributed to by the Br-4$p$ orbitals and the conduction band minimum by the Pb-6$p$ orbitals,[30] the non-linear pressure behavior up to 4.6 GPa has been explained in our previous study.[4] In particular, the observed behavior is a consequence of different distortions of the inorganic framework, with the decrease of the bond distance of Pb-Br favoring a band-gap decrease and the decrease of Pb-Br-Pb angle favoring a band-gap increase.[4] The first phenomenon dominates the pressure dependence of the band-gap in the low-pressure phase where the Pb-Br-Pb angle is constrained by symmetry to be 180º. The second phenomenon dominates the behavior from 0.9 to 4.6 GPa. In this pressure region the Pb-Br-Pb angle first changes from 180º to 165º at the first phase transition and then decreases upon further compression causing the observed opening of the band gap. This interpretation might also explain the 0.27 eV band gap opening at the phase transition at 4.6 GPa. It was recently shown that the distortions of geometry of the chains of PbI$_6$ octahedral in hybrid lead iodate perovskite could increase the band-gap energy from 2.10 eV to 2.55 eV. Confirmation of this hypothesis would require single-crystal XRD experiments up to 10 GPa.

## 4. Conclusions

In conclusion, in this work, we have reported the results of powder X-ray diffraction and optical-absorption experiments performed on MAPbBr$_3$ perovskite under high-pressure conditions. The two techniques, and visual observations, show that MAPbBr$_3$ does not become amorphous under compression up to at least 10.6 GPa. Instead of amorphization, a transition to a crystalline phase has been found at 4.6 GPa. The new phase is proposed to be orthorhombic with space group *Pmn*2$_1$. The phase transition to this new phase induces an abrupt increase of the band-gap energy. Our results contradict previous studies[5-7] but are in agreement with studies on MAPbI$_3$ which also ruled out amorphization[10] and found a transition to a crystalline phase at pressures where amorphization was previously reported, showing that the new high-pressure crystalline phase is stable up to 6 GPa.[10] Interestingly in both MAPbBr$_3$ and MAPbI$_3$ the structural phase transitions, which occur instead of the expected amorphization, trigger abrupt blue-shifts of the band-gap energy. The qualitative agreement between studies in MAPbBr$_3$ and MAPbI$_3$ suggest that the existence of new high-pressure phase, which could be disordered, and has a wider band gap, could be a common phenomenon in hybrid lead halide perovskites. In fact, this will not be the first case where the absence of a previously alleged pressure-induced amorphization has been reported.[31] Therefore, the present work shows that the disappearance of the XRD signal above a given pressure is not sufficient to be taken as the sole criterion for amorphization. Methods such as pair correlation distribution functions are needed to confirm the existence of amorphous compounds.[32] One of the possible reasons for pressure-induced amorphization in previous studies on MAPbBr$_3$ and MAPbI$_3$ could be the fact that experimental data may be affected by a large degree of sample non-hydrostaticity.[33] However, this hypothesis is ruled out by the fact that our study and the study on MAPbI$_3$[10] were performed under similar quasi-hydrostatic conditions as experiments where amorphization was reported. The reason for the previously reported amorphization could be related to other issues like different sample preparation methods or sample manipulation to load them into the diamond-anvil cell, which include methods from cutting samples into smaller pieces to mechanical grinding, which might introduce structural defects in a soft material like hybrid perovskites. Another possibility is the loading of large amounts of sample in the pressure chamber favoring sample bridging between the diamonds, which could induce large deviatoric stresses in the samples.[34]

**Authors Information**

**Corresponding Author**
Daniel Errandonea; ORCID: 0000-0003-0189-4221; Email: daniel.errandonea@uv.es

**Author Contributions**
A. L. and R. T. experiments, data analysis, writing-original draft, C. P. XRD experiments, I. F.-G., R. A., and P. B. synthesis of samples, D.E. conceptualization, data analysis, writing-original draft. All authors contributed to the discussion, writing, and proof-reading of the manuscript and have given approval to the final version of the manuscript.

**Notes**

The authors declare no competing financial interest.

**ACKNOWLEDGMENTS**

This work was supported by the Generalitat Valenciana under grants PROMETEO CIPROM/2021/075-GREENMAT and MFA/2022/007, and by the Spanish Ministerio de Ciencia e Innovación, Agencia Estatal de Investigación, and the European Union (MCIN/AEI/10.13039/501100011033) under grants PID2019–106383GB-41/44, RED2018–102612-T (MALTA Consolider Team), and Retos de la Sociedad Project Nirvana PID2020-119628RB-C31. This research is part of the Advanced Materials Programme supported by MCIN with funding from European Union NextGenerationEU (PRTR-C17.I1). R.T., D.E., and P.P.B. thank the Generalitat Valenciana for the postdoctoral fellowship no. CIAPOS/2021/20 and CIDEXG/2022/34. PXRD experiments were performed at the MSPD-BL04 beamline of ALBA Synchrotron (experiment no. 2021085271). We thank anonymous reviewers for constructive comments to the first version of this manuscript.